\documentclass[11pt]{article}

\usepackage{fullpage}
\usepackage{soul}
\usepackage[utf8]{inputenc}
\usepackage[small]{caption}
\usepackage{graphicx}
\usepackage{amsmath}
\usepackage{booktabs}
\usepackage{amsthm}
\usepackage{amssymb}
\usepackage{pifont}
\usepackage{bm,multirow}
\usepackage{color}
\usepackage{xspace}
\usepackage{tikz}
\usepackage[hidelinks]{hyperref}
\usepackage{cleveref}
\usepackage{tabularx}
\usepackage{pbox}
\usepackage{framed}
\usepackage{enumerate}
\usepackage{thm-restate}
\usepackage{nicematrix}
\usepackage{algorithm}
\usepackage{algorithmic}
\usepackage[most]{tcolorbox}
\usepackage{todonotes}
\usepackage[T1]{fontenc}

\usepackage{natbib}
\usepackage{babel}
\usepackage{subcaption}
\usetikzlibrary{arrows.meta}
\usetikzlibrary{positioning}
\renewcommand{\emptyset}{\varnothing}

\renewcommand{\emptyset}{\varnothing}

\newtheorem{theorem}{Theorem}[section]
\newtheorem{corollary}[theorem]{Corollary}

\newtheorem{lemma}[theorem]{Lemma}

\theoremstyle{definition}
\newtheorem{definition}[theorem]{Definition}

\newcommand*{\innerproofname}{Proof}

\allowdisplaybreaks

\usepackage{authblk}

\title{\bf Temporal Fair Division of Indivisible Mixed Manna: Tractable Settings}

\author[1]{Kui-Wang Choi} 
\author[1]{Minming Li}
\author[2]{Nicholas Teh}

\affil[1]{City University of Hong Kong, Hong Kong SAR}
\affil[2]{University of Oxford, United Kingdom}

\date{\vspace{-1cm}}

\begin{document}

\maketitle

\begin{abstract}
    We study temporal fair division of indivisible mixed manna. Items arrive over time and must be allocated irrevocably; an item may be a good for some agents, a chore for others, and neutral for the rest. We require the cumulative allocation after every round to be envy-free up to one item (TEF1). Although deciding whether a TEF1 allocation exists is NP-hard even for goods, we identify several tractable settings. First, with at most $k$ item types, an online cyclic rule guarantees EF$\lceil k/2\rceil$ after every item arrival. Thus, every instance with at most two types admits an online TEF1 allocation; moreover, when the numbers of agents and types are fixed, TEF1 existence can be decided in polynomial time. Second, under agreement after agent-specific scaling, provided that the scaling factors are known before arrivals begin, an online rule produces an allocation that is EF1 and Pareto optimal after every item arrival. Third, for a two-part arrival sequence with common rankings, we give a rule that is EF1 after every item arrival. Fourth, when the number of agents is fixed and values are bounded integers, we give an exact pseudo-polynomial algorithm for deciding TEF1 existence. Finally, for goods, every TEF1 allocation gives each agent at least $1/n$ of her maximin share after every round; this factor is tight even for identical valuations and two rounds. Deciding whether an exact temporal maximin-share allocation exists is NP-hard for both goods and chores, even with identical valuations, two agents, and two rounds.
\end{abstract}

\section{Introduction}

Fair division studies how indivisible resources and tasks should be
assigned among agents with different preferences
\citep{brams1996,Moulin03}. In a mixed manna instance, the
same item can be desirable to one agent, undesirable to another, and neutral to a third. A commonly studied desideratum is \emph{envy-freeness}, which requires every agent to value her own bundle at least as much as every other agent's bundle, but this may be impossible with indivisible items. For example, a single desirable item cannot be divided between two agents. A standard relaxation is \emph{envy-freeness up to one item} (EF1)
\citep{lipton2004}. For mixed manna, an allocation is EF1 if any
envy of agent $i$ toward agent $j$ can be eliminated in one of
two ways: remove from $i$'s bundle one item that $i$ values
nonpositively, or remove from $j$'s bundle one item that $i$
values nonnegatively \citep{aziz2022goodschores}.

Many allocation problems unfold over time. For example, teaching duties, on-call shifts, service requests, or time slots may become available on different days, and each assignment may require an immediate owner. In such settings, evaluating fairness only at the end can leave agents with little value or a heavy burden for many rounds, even if later assignments improve their final outcomes. We therefore evaluate the cumulative allocation after every round.

\emph{Temporal fair division} formalizes this setting and has received recent attention \citep{he2019,cookson2024,elkind2025temporalfd,choi2026,choi2026structured}.
An allocation is \emph{temporal EF1} (TEF1) if the cumulative allocation is EF1 after every round
\citep{elkind2025temporalfd}.%
\footnote{\emph{Temporal} and \emph{online} are different requirements. In a temporal problem, the full arrival sequence may be known before allocation begins, but assignments remain irrevocable. An online rule must assign the current item without using the identities, order, or values of later items; it may use parameters supplied before arrivals begin.}
However, TEF1 allocations need not exist. Even for goods, deciding whether a TEF1 allocation exists is NP-hard, and a related hardness result is
known for chores \citep{elkind2025temporalfd}. Existing positive results cover several restricted goods and chores settings.\footnote{The full version
of \citet{elkind2025temporalfd} gives a polynomial-time algorithm for mixed manna with two agents.}

Moving from goods-only or chores-only instances to mixed manna introduces an additional difficulty. Assigning an item to an agent may increase the value of the recipient's bundle for some agents but decrease it for others. Moreover, allocating separately the items that all agents value nonnegatively and those that all agents value nonpositively is insufficient: the two partial allocations may each be EF1 while their union is not. An algorithm for mixed manna must therefore evaluate the cumulative allocation as a whole.

Beyond the existence of TEF1 allocations, we ask when TEF1 can be achieved together with \emph{Pareto optimality}. An allocation is Pareto optimal if no alternative allocation makes every agent at least as well off and at least one agent strictly better off. We further study \emph{temporal maximin share} (TMMS) fairness. For goods, agent $i$'s \emph{maximin share} (MMS) is the largest value she can guarantee by partitioning the available items into n bundles and then receiving the bundle she values least \citep{budish2011}. TMMS requires this guarantee after every round. TEF1 and MMS assess fairness differently: TEF1 compares the bundles in the chosen allocation, whereas MMS compares an agent's value with what she can guarantee through an $n$-way partition.

These observations raise the following questions:
\begin{quote}
    \emph{Under what conditions does a TEF1 allocation of mixed manna always exist, and when can one be found efficiently? When can TEF1 be achieved together with Pareto optimality? What guarantees are possible for temporal MMS?}
\end{quote}

For TEF1, we study four conditions: a small number of item types, agreement after scaling each agent's values by a fixed positive factor, common rankings in a two-part arrival sequence, and bounded integer values. For temporal MMS, we determine the guarantee implied by TEF1 and study the complexity of deciding whether an exact TMMS allocation exists.

\subsection{Our Results}
We give three allocation rules for mixed manna, two algorithms for
deciding TEF1 existence, and two results on temporal maximin share
fairness. Unless stated otherwise, each allocation rule guarantee holds after every item arrival and therefore also at every round end. The first two rules are online: they assign the current item without using later arrivals.

\paragraph{Few item types.}
Two items have the same type if every agent assigns them the same value.
For an instance with at most $k$ types, we give an online rule that uses a
fixed cycle for each type and is EF$\lceil k/2\rceil$ after
every item arrival. Here EF$\ell$ means envy-free up to
$\ell$ items. In particular, every instance with at most two types has
an online TEF1 allocation. Even within a single type, some agents may value the items positively, others negatively, and others at zero. This extends the earlier two-type result for
instances containing only goods or only chores (Elkind et al. 2025). For
fixed $n$ and $k$, we also give a polynomial-time dynamic program that decides whether a TEF1 allocation exists and returns one if it does.

\paragraph{Agreement after agent-specific scaling.}
Assume there are fixed positive scales $c_1,\dots,c_n$. For an item $o$,
agent $i$'s normalized value is $v_i(o)/c_i$. We require all positive
normalized values for the same item to be equal. We also require all
normalized values to be equal when every agent values the item
negatively. Given the scales, our online rule uses each agent's current
normalized bundle value. It assigns an item that some agent values positively to a positive-valuing agent with minimum current normalized bundle value. If no agent values the item positively but some agent values it at zero, the rule assigns the item to a zero-valuing agent. If all agents value the item negatively, the rule assigns it to an agent with maximum current normalized bundle value. The
allocation is EF1 and Pareto optimal after every item arrival. If $A_i$
denotes agent $i$'s bundle, the rule also maximizes $\sum_{i\in N}\frac{v_i(A_i)}{c_i}$
among all allocations of the items that have arrived. For rational
valuations, valid scales can be found in polynomial time when they exist.
The class includes scaled ternary valuations,
$v_i(o)\in\{-c_i,0,c_i\}$, and proportional valuations,
$v_i(o)=c_iw(o)$.

\paragraph{A two-part sequence with common rankings.}
We next consider a sequence with a first part $C$ and a second part $G$.
Every agent assigns nonpositive value to every item in $C$ and weakly prefers earlier items of $C$. Every agent in a fixed nonempty set $P_G$ assigns nonnegative value to every item in $G$ and weakly prefers earlier items of $G$, whereas every agent outside $P_G$ assigns nonpositive value to every item in $G$. We allocate $C$ cyclically among all agents and allocate $G$ cyclically among the agents in $P_G$ in the reverse label order. The resulting allocation is EF1 after every item
arrival, even when agents in $P_G$ assign different numerical values to
the same item. The split between $C$ and $G$ and the set $P_G$ must be
known before the first assignment, so this rule is not online.

\paragraph{Bounded integer values.}
Fix $n$ and suppose every item value is an integer in $[-U,U]$. We give
an exact dynamic program whose running time is polynomial in the input size and in $U$. It decides whether a TEF1 allocation exists and returns
one when it does. Thus the running time is pseudo-polynomial when $U$ is
written in binary. The algorithm accommodates any number of item types and any pattern of positive, zero, and negative values. Among all TEF1
allocations, the algorithm can optimize any polynomial-time objective
determined by the final value matrix $(v_i(A_j))_{i,j\in N}$, whose
$(i,j)$-entry is agent $i$'s value for agent $j$'s final bundle. Examples
include $\sum_i v_i(A_i)$ and $\min_i v_i(A_i)$.

\paragraph{Temporal maximin share fairness.}
An allocation is $\alpha$-TMMS for goods if every agent receives at least
an $\alpha$ fraction of her MMS after every round. For additive goods,
every TEF1 allocation is $1/n$-TMMS. This factor cannot be improved: for
every $n\ge 2$, there is a two-round goods instance with identical
valuations that has a TEF1 allocation but has no $\alpha$-TMMS allocation
for any $\alpha\in(1/n,1]$. We also prove that deciding whether an exact
TMMS allocation exists is NP-hard for two agents and two rounds, even
with identical valuations, for both goods and chores.

\subsection{Related Work}

\paragraph{Temporal fair division.}
\citet{he2019} require EF1 after every round and study
how many earlier assignments must be changed. In their informed
setting with two agents, EF1 can be maintained without changing an
earlier assignment. \citet{elkind2025temporalfd} study
immediate and irrevocable assignment when the full arrival sequence is
known. For goods and chores, they give polynomial-time results for two
agents, two item types, generalized binary values, and unimodal
preferences, and they prove NP-hardness in general. They also show that
TEF1 and Pareto optimality may be incompatible. Their full version
further proves that, for two agents, a TEF1 allocation of arbitrary
mixed manna always exists and can be found in polynomial time. Our item-type result instead handles any number of agents and gives an online TEF1 rule for at most two item types, even when a type is valued positively by some agents, negatively by others, and at zero by the rest.
Our result in Section~\ref{sec:scaled-agreement} identifies a class of mixed manna valuations for which an online rule is EF1 and Pareto optimal after every item arrival.
\citet{cookson2024} study goods and combine fairness within
each round with fairness after all rounds or after every prefix of
rounds. \citet{choi2026} study a stronger temporal envy-freeness requirement and temporal maximin shares for goods, and they also permit limited delays in assigning items. \citet{Goldberg2026} study the complementary objective of minimizing cumulative envy over an allocation sequence.

\paragraph{Mixed manna and Pareto optimality.}
Here, a \emph{non-temporal} allocation is one for which only the final
allocation is evaluated; no fairness condition is imposed during an
arrival sequence.
\citet{bogomolnaia2017mixed} introduce mixed manna for divisible items. 
For indivisible mixed manna with additive values, \citet{aziz2022goodschores} define EF1, prove that an EF1 allocation always exists,
and obtain EF1 together with Pareto optimality for two agents. Their
double round-robin rule has two consecutive cyclic phases in opposite
orders: it first allocates items that every agent values nonpositively
and then allocates the remaining items. This rule is the closest non-temporal analogue of our two-part common-ranking rule, which keeps
the given arrival order and is EF1 after every item arrival. \citet{bhaskar2021mixed} prove EF1 existence in a more general
mixed manna model. Our EF$\ell$ definition follows \citet{berczi2024relaxations}. \citet{liu2024survey} survey the mixed manna literature, and \citet{garg2024unequal} study unequal entitlements.

\citet{aleksandrovWalsh2020} give a non-temporal allocation rule whose local recipient choice coincides with that of our Algorithm~2 when all scales equal $1$. Their algorithm first reorders the items by value magnitude and then produces an allocation that is envy-free up to any item (EFX) and Pareto optimal after every step of that chosen order. Our rule keeps the actual arrival order, proves EF1 and Pareto optimality after every item arrival, and permits
agent-specific positive scales. \citet{livanos2022mixed}
study restricted mixed goods, where agents who value an item positively assign it the same positive value. Our condition permits agent-specific scales and requires equal normalized values for an item
valued negatively by every agent.

For general non-temporal mixed manna, \citet{barman2026mixed} obtain Pareto optimality with EFR-$(n-1)$, whose envy test may reassign at most $n-1$ items.
\citet{barman2025introspective} obtain Pareto optimality with IEF1, which lets an agent add one item to or remove one item from her own bundle when checking envy. \citet{aziz2026bobw} prove that a randomized non-temporal allocation can be envy-free in expectation while every realized allocation is EF1.
Our result in Section~\ref{sec:scaled-agreement} instead uses the standard
mixed manna EF1 definition, is deterministic, never changes an assigned
item, and gives Pareto optimality after every item arrival.

\paragraph{Maximin shares.}
MMS was introduced by \citet{budish2011}. Exact MMS
may fail even for additive goods, which led to approximation
results~\citep{kurokawa2018fairenough}. MMS for non-temporal mixed manna is studied by \citet{kulkarni2021ptas}, and MMS together with Pareto optimality by
\citet{kulkarni2021mmspo}. \citet{amanatidis2018comparing} prove that every EF1 goods
allocation gives each agent at least $1/n$ of her MMS. \citet{choi2026} show that exact TMMS may fail in temporal
goods instances. We apply the non-temporal EF1 implication separately at each round, prove that $1/n$ is the best universal TMMS factor even for identical values and two rounds, and prove NP-hardness of exact TMMS existence for both goods and chores with two agents and two
rounds. \citet{Lim2026repeated} maximize the minimum realized cumulative utility across agents in a repeated matching model. Their objective compares the utilities obtained by different agents. MMS is different: each agent's MMS is computed separately from that agent's own valuation by considering all $n$-way partitions of the items.

\section{Preliminaries} \label{sec:preliminaries}

For a positive integer $z$, let $[z]:=\{1,\dots,z\}$. An instance of \emph{temporal fair division} is a tuple $I=\langle N,T,(O_t)_{t\in[T]},v\rangle$.
Here $N=[n]$ is the set of agents, $T$ is the number of rounds, and $O_t$ is the set of items arriving in round $t$. The sets $O_1,\dots,O_T$ are pairwise disjoint. Let $O:=\bigcup_{t=1}^T O_t$, $O_{\le t}:=\bigcup_{\tau=1}^t O_\tau$, $m:=|O|$.

The valuation profile is $v=(v_1,\dots,v_n)$. For each agent $i$, the number $v_i(o)\in\mathbb{R}$ is the value of item $o$ to agent $i$. We extend $v_i$ additively to sets of items: $v_i(S):=\sum_{o\in S}v_i(o)$ for every $S \subseteq O$.
When all agents have the same valuation function, we write $v$ instead of $v_i$.

For $S\subseteq O$, let $\Pi_n(S)$ be the set of tuples
$(A_1,\dots,A_n)$ such that the sets $A_1,\dots,A_n$ are pairwise
disjoint and their union is $S$; empty sets are allowed. An
\emph{allocation} of $S$ is a tuple $A=(A_1,\dots,A_n)\in\Pi_n(S)$,
where $A_i$ is agent $i$'s bundle. A \emph{full allocation} is an
allocation of $O$. For a full allocation $A$ and a round $t$, let $A_i^t:=A_i\cap O_{\le t}$ and $A^t:=(A_1^t,\dots,A_n^t)$.
Thus $A^t$ is the cumulative allocation after round $t$, and $A=A^T$.

For $S\subseteq O$, an allocation $A'\in\Pi_n(S)$ \emph{Pareto dominates} $A\in\Pi_n(S)$ if $v_i(A'_i)\ge v_i(A_i)$ for every $i \in N$, with a strict inequality for at least one agent. An allocation is \emph{Pareto optimal} if no other allocation Pareto dominates it.

When discussing item-by-item procedures, fix an order within each round and write $o_1,\dots,o_m$ for the resulting order, with every item of round $t$ appearing before every item of round $t+1$. For $s\in\{0,\dots,m\}$, let $Q_s:=\{o_1,\dots,o_s\}$ and $Q_0:=\varnothing$.
Thus $Q_s$ is the set of the first $s$ items in this order. Let $b_t:=|O_{\le t}|=\sum_{\rho=1}^t|O_\rho|$ be the number of items that have arrived by the end of round $t$.
Temporal fairness is required for the allocations of
$Q_{b_1},\dots,Q_{b_T}$. Some of our rules satisfy the stronger condition of fairness after every item arrival.

The temporal model may reveal the full instance before allocation begins.
We call a rule \emph{online} if the recipient of item $o_s$ depends only
on the values of $o_1,\dots,o_s$, the assignments already made, and
parameters fixed before the arrival process begins. It may not use the
number, identities, order, or values of later items. Every assignment is
irrevocable. In Section~\ref{sec:scaled-agreement}, the scales are treated
as fixed parameters supplied before the arrivals; the separate procedure
for finding scales examines the full valuation profile and is not part of
the online rule.

Agent $i$ \emph{envies} agent $j$ under allocation $A$ if $v_i(A_i)<v_i(A_j)$. The allocation is \emph{envy-free} if no such ordered pair exists.

We consider mixed manna, so an item may have positive, negative, or zero
value. For agent $i$, item $o$ is a \emph{weak good} if $v_i(o)\ge 0$
and a \emph{weak chore} if $v_i(o)\le 0$. A zero-valued item is both a
weak good and a weak chore. We say that an instance \emph{contains only goods} if $v_i(o)\ge 0$ for every $i$ and $o$, and \emph{contains only chores} if $v_i(o)\le 0$ for every $i$ and $o$.

\begin{definition}[EF$\ell$ for mixed manna]
\label{def:efl}
Fix $S\subseteq O$ and $\ell\in\mathbb{Z}_{\ge 0}$. An allocation
$A=(A_1,\dots,A_n)\in\Pi_n(S)$ is \emph{envy-free up to $\ell$ items}
(EF$\ell$) if, for every ordered pair of distinct agents
$i,j\in N$, there are sets $B_{ij}\subseteq A_i$ and
$G_{ij}\subseteq A_j$ such that $|B_{ij}|+|G_{ij}|\le \ell$, every item in $B_{ij}$ is a weak chore for $i$, every item in $G_{ij}$ is a weak good for $i$, and $v_i(A_i\setminus B_{ij})\ge v_i(A_j\setminus G_{ij})$.
\end{definition}

For EF1, at most one item is deleted in total. Thus, when comparing $i$ with $j$, one may delete either one item from $i$'s bundle that $i$ views as a weak chore or one item from $j$'s bundle that $i$ views as a weak good, but not one item from each bundle.

\begin{definition}[Temporal EF$\ell$]
\label{def:tefl}
    A full allocation $A\in\Pi_n(O)$ is \emph{temporal
    EF$\ell$} (TEF$\ell$) if $A^t$ is
    EF$\ell$ for every $t\in[T]$. A partial allocation through round $t$ is TEF$\ell$ if the cumulative allocation is EF$\ell$ after each of rounds $1,\dots,t$.
\end{definition}

Unless stated otherwise, EF$\ell$ and TEF$\ell$ are as in
Definitions~\ref{def:efl} and~\ref{def:tefl}, respectively. For a goods instance, one may take $B_{ij}=\varnothing$; for a chores instance, one may take $G_{ij}=\varnothing$.

\section{Few Item Types}
 \label{sec:limited-item-diversity}
We first assume that only a small number of distinct item-value vectors occur. Two items $o,o'\in O$ have the same \emph{type} if $v_i(o)=v_i(o')\quad\text{for every }i\in N$.
Suppose the instance has $q\le k$ types, denoted by $S_1,\dots,S_q$. For every agent $i$ and type $r$, let $w_{i,r}$ be the common value that agent $i$ assigns to every item in $S_r$.

The type assumption lets us describe an allocation by the number of items
of each type assigned to each agent. \citet{elkind2025temporalfd} prove TEF1 existence for two types in goods-only and chores-only instances. In our mixed manna setting, even a single type may be valued positively by some agents, negatively by others, and at zero by the rest.

For each type $r$, define $P_r:=\{i\in N:w_{i,r}>0\}$, $Z_r:=\{i\in N:w_{i,r}=0\}$, and $C_r:=\{i\in N:w_{i,r}<0\}$.
The three sets partition the agents according to their value for type $r$.

\begin{algorithm}[t]
\caption{Cyclic allocation with few item types}
\label{alg:few-types}
\begin{algorithmic}[1]
\STATE $A_i\gets\varnothing$ for every $i\in N$; $F\gets\varnothing$;
$R\gets\varnothing$
\FOR{each arriving item $o$ in the fixed order}
    \STATE let $r$ be the type of $o$
    \IF{$r\notin F\cup R$}
        \IF{$|F|\le |R|$}
            \STATE $F\gets F\cup\{r\}$
            \STATE let $\sigma_r$ list $P_r$ increasingly, then $Z_r$
            increasingly, then $C_r$ decreasingly
        \ELSE
            \STATE $R\gets R\cup\{r\}$
            \STATE let $\sigma_r$ list $P_r$ decreasingly, then $Z_r$
            increasingly, then $C_r$ increasingly
        \ENDIF
        \STATE $z_r\gets 1$
    \ENDIF
    \STATE $h\gets \sigma_r(z_r)$
    \STATE $A_h\gets A_h\cup\{o\}$
    \STATE $z_r\gets 1+(z_r\bmod n)$
\ENDFOR
\RETURN $A$
\end{algorithmic}
\end{algorithm}

When a type first appears, Algorithm~\ref{alg:few-types} places it in one
of two sets, $F$ or $R$, and fixes an ordered list
$\sigma_r=(\sigma_r(1),\dots,\sigma_r(n))$ of the agents. Successive items of type $r$ are assigned to
$\sigma_r(1),\dots,\sigma_r(n),\sigma_r(1),\dots$ in that order. For a type in $F$, the list contains the agents in $P_r$ by increasing label, then the agents in $Z_r$ by increasing label, and then the agents in $C_r$ by decreasing label. For a type in $R$, the positive and negative groups use the opposite label directions. A new type is placed in the smaller of $F$ and $R$, with ties broken in favor of $F$. Therefore, after $q'$ types have appeared, $\max\{|F|,|R|\}\le \left\lceil\frac{q'}{2}\right\rceil$.

For any prefix $Q_s$, let $x_{i,r}$ be the number of type-$r$ items assigned to agent $i$. For each type, the counts of any two agents differ by at most one. The chosen orders also imply $w_{i,r}(x_{i,r}-x_{j,r})\ge 0$ when $i<j$ and $r\in F$, and when $i>j$ and $r\in R$. Hence, for
$i<j$, only types in $R$ can lower the difference
$v_i(A_i)-v_i(A_j)$; for $i>j$, only types in $F$ can do so. Removing at
most one item for each such type removes the envy.

\begin{theorem}\label{thm:few-types}
    For an instance with $q$ item types,
Algorithm~\ref{alg:few-types} is online and produces an allocation that is EF$\lceil q/2\rceil$ after every item arrival. Consequently,
every instance with at most $k$ item types has an online allocation that
is EF$\lceil k/2\rceil$ after every item arrival.
\end{theorem}

\paragraph{For $q=3$, the allocation produced by Algorithm~\ref{alg:few-types} may require two deletions.}
Let $n=q=3$. One item of each of the following three types arrives in the
displayed order, where each coordinate is the value of the corresponding
agent: $(1,1,1),\quad (0,1,-1),\quad (-1,-1,-1)$.
The tie rule in Algorithm~\ref{alg:few-types} places the first type in
$F$, the second in $R$, and the third in $F$. The three ordered lists are $(1,2,3),\quad (2,1,3),\quad (3,2,1)$,
respectively. Hence the three items are assigned to agents $1$, $2$, and
$3$.
Agent 3 values her own bundle at $-1$ and agent 1's bundle at $1$. After deleting agent 1's item from agent 1's bundle, agent 3 values the two compared bundles at $-1$ and $0$; after deleting agent 3's item from her own bundle, she values them at $0$ and $1$. Neither permitted single deletion eliminates the envy, so the allocation is not EF1. Deleting both items gives
value $0$ on both sides, so the allocation is EF2.

For $k\le 2$, Theorem~\ref{thm:few-types} gives an online TEF1
allocation. This extends the two-type goods and chores result of \citet{elkind2025temporalfd}, because the same type may be valued positively by some agents, negatively by others, and at zero by the rest.

The type-count representation also yields an exact decision algorithm. Fix the item
order $o_1,\dots,o_m$. For $\tau\in\{0,\dots,m\}$ and type $r$, let $p_r(\tau):=|\{a\in[\tau]:o_a\in S_r\}|$
be the number of type-$r$ items among the first $\tau$ items. A \emph{state} after $\tau$ items is a count matrix
$x=(x_{i,r})_{i\in N,r\in[q]}$ satisfying $\sum_{i\in N}x_{i,r}=p_r(\tau)$ for every $r \in [q]$.
The entry $x_{i,r}$ is the number of type-$r$ items assigned to agent
$i$.

For distinct agents $i$ and $j$, define $\Delta_{ij}(x):=\sum_{r=1}^q(x_{j,r}-x_{i,r})w_{i,r}$.
This is the amount by which agent $i$ values agent $j$'s bundle above her
own; a positive value means that $i$ envies $j$. Also define $\eta_{ij}(x):=
\max(
\{w_{i,r}:x_{j,r}>0,\ w_{i,r}>0\}
\cup
\{-w_{i,r}:x_{i,r}>0,\ w_{i,r}<0\}
\cup\{0\}
)$.
The quantity $\eta_{ij}(x)$ is the largest decrease in
$\Delta_{ij}(x)$ that one deletion allowed by EF1 can produce. Therefore $x$ describes an EF1 allocation if and only if $\Delta_{ij}(x)\le \eta_{ij}(x)$ for every $i \neq j$.

Let $B:=\{b_t:t\in[T]\}$ be the set of round-end indices. Let $D_\tau$
be the set of states reachable after the first $\tau$ assignments that
satisfy EF1 at every round end already completed. For each stored state,
the algorithm saves one previous state and the recipient of the last item;
these saved choices allow the allocation to be reconstructed at the end.
The dynamic program tries every possible recipient for
$o_{\tau+1}$, keeps one copy of each resulting count matrix,
and applies the EF1 test only when $\tau+1\in B$.
Full pseudocode appears in
Appendix~\ref{app:few-types-dp}.

\begin{theorem} \label{thm:few-types-dp}
For fixed $n$ and $k$, TEF1 existence for instances with at most $k$ item
types can be decided in polynomial time. When a TEF1 allocation exists, the dynamic program returns one.
\end{theorem}

The implementation needs to store only the counts $x_{i,r}$ for $i\in[n-1]$, because the final agent's count is $x_{n,r}=p_r(\tau)-\sum_{i=1}^{n-1}x_{i,r}$.
Hence the number of states after any $\tau$ is at most $\prod_{r=1}^q(p_r(\tau)+1)^{n-1}
\le (m+1)^{q(n-1)}
\le (m+1)^{k(n-1)}$.
Testing whether a state is EF1 takes $O(n^2q)$ time. Thus the running time is polynomial when $n$ and $k$ are fixed.

\section{Agreement after Agent-Specific Scaling}
\label{sec:scaled-agreement}
The previous section assumes a small number of item types. We now allow arbitrarily many distinct item-value vectors and ask when an online rule can guarantee both EF1 and Pareto optimality.\footnote{The combination of fairness and efficiency guarantees has been well-studied in the offline fair division literature \citep{BarmanKV18,CaragiannisKMPPSW19,Lim2026EFXcost,Mahara2026,Teh2026,Teh2026EF1PO}.} Here, agreement requires equality of normalized numerical values; it is stronger than agreement only on signs or rankings.

Allocating the positive and negative items separately is not sufficient.
For example, consider two agents with the same valuation, with
$v(g)=2$ and $v(c)=-2$. Assign $g$ to agent 2 and $c$ to agent 1. The
allocation of $\{g\}$ is EF1, and the allocation of $\{c\}$ is EF1. Their union is not EF1: from agent 1's viewpoint, deleting $g$ from agent 2's bundle leaves values $-2$ and $0$, whereas deleting $c$ from agent 1's own bundle leaves values $0$ and $2$. 
Thus the rule must compare the agents' complete current bundles.

\begin{definition}[Agreement after scaling]
\label{def:scaled-agreement}
A valuation profile $v$ satisfies \emph{agreement after scaling} if there
are positive numbers $c_1,\dots,c_n$ such that the normalized values $a_i(o):=\frac{v_i(o)}{c_i}$ satisfy the following condition for every item $o$. Let $P_o:=\{i\in N:a_i(o)>0\}$.
Exactly one of the following cases applies:
\begin{enumerate}[(i)]
    \item $P_o\ne\varnothing$, and there is a number $p_o>0$ such that
    $a_i(o)=p_o$ for every $i\in P_o$;
    \item $P_o=\varnothing$ and $\max_{i\in N}a_i(o)=0$;
    \item there is a number $q_o<0$ such that $a_i(o)=q_o$ for every
    $i\in N$.
\end{enumerate}
For $S\subseteq O$, denote $a_i(S):=\sum_{o\in S}a_i(o)=\frac{v_i(S)}{c_i}$.
A vector $(c_1,\dots,c_n)$ satisfying these conditions is called a
\emph{valid scale vector}.
\end{definition}

Thus, if at least one agent values an item positively, every agent who
values it positively has exactly the same normalized value. Agents who do
not value that item positively may assign it zero or a negative value. If
no agent values the item positively, then either at least one normalized
value is zero, or all normalized values are equal and negative.

For rational valuations, a valid scale vector can be found in
$O(mn^2)$ arithmetic operations when one exists.
For each item, we impose
$c_i/c_j=v_i(o)/v_j(o)$ for every pair of agents who value
the item positively. If every agent values the item negatively,
we impose the same equation for every pair of agents.
These equations define a graph on the agents. Starting from one agent in each connected component, we determine the remaining scales from these ratios and reject any conflicting equation. The full procedure appears in
Appendix~\ref{app:find-scales}.

The online allocation theorem below assumes that a valid
scale vector is supplied before the arrivals begin.

For a partial allocation $A$, define agent $i$'s current normalized
bundle value by $L_i:=a_i(A_i)=\frac{v_i(A_i)}{c_i}$.
The value $L_i$ may be positive, zero, or negative.

\begin{algorithm}[t]
\caption{Online allocation under agreement after scaling}
\label{alg:scaled-allocation}
\begin{algorithmic}[1]
\STATE $A_i\gets\varnothing$ and $L_i\gets0$ for every $i\in N$
\FOR{each arriving item $o$ in the fixed order}
    \STATE $P_o\gets\{i\in N:a_i(o)>0\}$
    \IF{$P_o\ne\varnothing$}
        \STATE choose the smallest-label agent
        $h\in\arg\min_{i\in P_o}L_i$
    \ELSIF{$\max_{i\in N}a_i(o)=0$}
        \STATE choose the smallest-label agent
        $h\in\{i\in N:a_i(o)=0\}$
    \ELSE
        \STATE choose the smallest-label agent
        $h\in\arg\max_{i\in N}L_i$
    \ENDIF
    \STATE $A_h\gets A_h\cup\{o\}$ and $L_h\gets L_h+a_h(o)$
\ENDFOR
\RETURN $A$
\end{algorithmic}
\end{algorithm}

In every case, the chosen recipient maximizes the normalized value of the
current item. Consequently, every assigned item $x\in A_j$ satisfies
$a_i(x)\le a_j(x)$ for every agent $i$, and hence $a_i(A_j)\le a_j(A_j)=L_j$.
The comparisons among the $L_i$ values in Algorithm~\ref{alg:scaled-allocation} control the two nontrivial cases: a
positively valued item goes to an agent in $P_o$ with minimum $L_i$, whereas
an item valued negatively by everyone goes to an agent with maximum $L_i$.
In the remaining case, the recipient values the item at zero and every other
agent values it nonpositively, so no comparison worsens. These observations
drive the EF1 proof.

\begin{theorem}
\label{thm:scaled-allocation}
    Given a valid scale vector, Algorithm~\ref{alg:scaled-allocation} is
    online, runs in $O(mn)$ time, and produces an allocation that is EF1 after
    every item arrival. For every $s\in\{0,\dots,m\}$, the allocation of
    $Q_s$ maximizes $\sum_{i\in N}\frac{v_i(A_i)}{c_i}$ over all allocations of $Q_s$. Consequently, the allocation is Pareto optimal after every item arrival.
\end{theorem}

\paragraph{Relation to \citet{aleksandrovWalsh2020}.}
When $c_i=1$ for every agent, Algorithm~\ref{alg:scaled-allocation}
chooses the same local recipient as Algorithm 2 of \citet{aleksandrovWalsh2020}. Their algorithm first reorders the items by the absolute value
of their maximum utility. Our rule keeps the given arrival order, and
Theorem~\ref{thm:scaled-allocation} proves EF1 and Pareto optimality after
every actual item arrival. The theorem also permits agent-specific
scales.

When $c_i=1$ and, for every item $o$, there is a nonzero number $\beta_o$
such that $v_i(o)\in\{0,\beta_o\}$ for every agent $i$, the same rule
specializes to the generalized binary goods and chores rules of \citet{elkind2025temporalfd}.\footnote{Generalized binary valuations (also commonly referred to as \emph{cost utilities} or \emph{restricted additive valuations} \citep{AkramiReSe22,CamachoFePe23}) generalizes the commonly-studied class of \emph{identical valuations} \citep{Choo2024,Elkind2024,plaut2018efx,Scarlett2023} and \emph{binary valuations} \citep{aleksandrov2015,Brandl2026binary,halpern2020binary,neoh2025efx,SuksompongTe22,WangWei2026}.}

\begin{corollary}
\label{cor:scaled-classes}
Each of the following valuation classes has an online allocation that is EF1 and Pareto optimal after every item arrival:
\begin{enumerate}[(i)]
    \item scaled ternary valuations: there are positive numbers
    $c_1,\dots,c_n$ such that $v_i(o)\in\{-c_i,0,c_i\}$ for every $i\in N$ and $o\in O$;
    \item proportional valuations: there are positive numbers
    $c_1,\dots,c_n$ and item weights $w(o)\in\mathbb{R}$ such that $v_i(o)=c_iw(o)$ for every $i\in N$ and $o\in O$.
\end{enumerate}
\end{corollary}

\section{Common Rankings in a Two-Part Sequence}
\label{sec:common-ranking}

Section~\ref{sec:scaled-agreement} requires equality of normalized numerical values. We now allow agents to assign different numerical values, but require them to share a common ranking of the relevant items. A common ranking in arrival order means that every relevant agent weakly prefers each earlier item to each later item.

The sequence consists of a first part $C$, followed by a second part $G$. Every agent views every item in $C$ as a weak chore. If $G\neq\emptyset$, every agent in a fixed nonempty set $P_G$ views every item in $G$ as a weak good, whereas every agent outside $P_G$ views every item in $G$ as a weak chore. The split between $C$ and $G$, the starting offset for the $C$-cycle, and the set $P_G$ are determined from the full sequence. The rule is therefore not online, although it is EF1 after every item arrival.

\begin{definition}[Two-part common-ranking sequence]
\label{def:common-ranking}
Fix the item order $o_1,\dots,o_m$. A \emph{two-part common-ranking
sequence} has a split index $b\in\{0,\dots,m\}$ and the two consecutive
parts $C:=\{o_1,\dots,o_b\}$ and $G:=\{o_{b+1},\dots,o_m\}$.
The following conditions hold.
\begin{enumerate}[(i)]
    \item For every agent $i$ and every $o\in C$, $v_i(o)\le0$.
    Moreover, for $1\le a<a'\le b$, $v_i(o_a)\ge v_i(o_{a'})$.
    Thus every agent weakly prefers an earlier item of $C$ to a later
    item of $C$.

    \item If $G\ne\varnothing$, there is a nonempty set $P_G\subseteq N$
    such that $v_i(o)\ge0$ for every $i\in P_G$ and $o\in G$, while
    $v_i(o)\le0$ for every $i\notin P_G$ and $o\in G$. Moreover, for
    every $i\in P_G$ and $b<a<a'\le m$, $v_i(o_a)\ge v_i(o_{a'})$.
\end{enumerate}
If $G=\varnothing$, no set $P_G$ is needed.
\end{definition}

A valid split and set $P_G$, when they exist, can be found
in $O(m^2n)$ time by trying all split indices.
For a fixed split with $G\neq\varnothing$, let
$P^+:=\{i\in N : v_i(o)>0\text{ for some }o\in G\}$.
If $P^+\neq\varnothing$, it is enough to test $P_G=P^+$.
If $P^+=\varnothing$, a valid nonempty $P_G$ exists exactly
when some agent assigns value zero to every item in $G$.
The case $G=\varnothing$ only requires condition~(i) of
Definition~\ref{def:common-ranking}. Full pseudocode appears in
Appendix~\ref{app:find-split}.

Choose $d\in\{0,\dots,n-1\}$ such that $d+|C|\equiv0\pmod n$.
The choice of $d$ shifts the starting agent so that the allocation of $C$
ends after a complete cycle. Equivalently, in the proof one may place $d$
zero-valued items before $C$, cycle through agents $1,2,\dots,n$, and
then ignore those added zero-valued items.

\begin{algorithm}[t]
\caption{Allocating a two-part common-ranking sequence}
\label{alg:common-ranking-allocation}
\begin{algorithmic}[1]
\STATE $A_i\gets\varnothing$ for every $i\in N$
\STATE choose $d\in\{0,\dots,n-1\}$ such that
$d+|C|\equiv0\pmod n$
\STATE $h\gets d+1$
\FOR{each item $o\in C$ in arrival order}
    \STATE $A_h\gets A_h\cup\{o\}$
    \STATE $h\gets 1+(h\bmod n)$
\ENDFOR
\IF{$G\ne\varnothing$}
    \STATE write $P_G=\{p_1<\cdots<p_g\}$ and set $u\gets g$
    \FOR{each item $o\in G$ in arrival order}
        \STATE $A_{p_u}\gets A_{p_u}\cup\{o\}$
        \STATE $u\gets u-1$; if $u=0$, set $u\gets g$
    \ENDFOR
\ENDIF
\RETURN $A$
\end{algorithmic}
\end{algorithm}

\begin{lemma}
\label{lem:chore-cycle}
Suppose $C$ satisfies condition (i) of
Definition~\ref{def:common-ranking}. Choose
$d\in\{0,\dots,n-1\}$ such that $d+|C|\equiv0\pmod n$, and allocate
$C$ cyclically starting with agent $d+1$. The allocation is EF1 after
every item arrival in $C$. After all items of $C$ are assigned, $v_i(A_i)\geq v_i(A_j)$ for every $i<j$. Conversely, for every $i<j$, any envy of $j$ toward $i$ can be removed by deleting one weak chore from $j$'s own bundle.
\end{lemma}

\begin{theorem}
\label{thm:common-ranking}
After a valid split and set $P_G$ are found as above,
Algorithm~\ref{alg:common-ranking-allocation} produces in polynomial time an
allocation that is EF1 after every item arrival. Consequently, every two-part
common-ranking sequence can be allocated in polynomial time with this
property.
\end{theorem}

The agreement-after-scaling class and the two-part
common-ranking class are incomparable; examples in both
directions appear in
Appendix~\ref{app:class-comparison}.

\section{Exact Algorithms for Bounded Integral Valuations}
\label{sec:integer-values}

The preceding sections use a small number of item types, equality of
normalized values, or a common ranking. We now allow any number of item
types and any pattern of positive, zero, and negative values. General
TEF1 existence is NP-hard even for goods \citep{elkind2025temporalfd}. We fix the
number of agents and assume that every item value is an integer in
$[-U,U]$.

For a partial allocation and distinct agents $i,j$, define $\Delta_{ij}:=v_i(A_j)-v_i(A_i)$, $p_{ij}:=\max(\{v_i(o):o\in A_j,\ v_i(o)>0\}\cup\{0\})$, and $q_i:=\max(\{-v_i(o):o\in A_i,\ v_i(o)<0\}\cup\{0\})$.
The quantity $\Delta_{ij}$ is the amount by which agent $i$ values
agent $j$'s bundle above her own. The quantity $p_{ij}$ is the largest
positive value that $i$ assigns to one item in $j$'s bundle. The quantity $q_i$ is the largest magnitude of a negative value that agent $i$ assigns to an item in her own bundle.

If $\Delta_{ij}\le0$, agent $i$ does not envy agent $j$. If $\Delta_{ij}>0$, a permitted EF1 deletion can reduce the envy gap by at most $p_{ij}$ or $q_i$. Thus, the current allocation is EF1 if and only if
\begin{equation}
\label{eq:integer-ef1-test}
\Delta_{ij}\le\max\{p_{ij},q_i\}
\quad\text{for every }i\ne j.
\end{equation}

\begin{theorem}
\label{thm:integer-dp}
Fix $n$ and let $U\in\mathbb{Z}_{\ge0}$. Suppose
$|v_i(o)|\le U$ for every agent $i$ and item $o$, and every value is an
integer. Then TEF1 existence can be decided in time polynomial in $m$, $T$,
and $U$, and a TEF1 allocation can be returned whenever one exists. When $U$
is written in binary, this is a pseudo-polynomial running time.

Among all TEF1 allocations, the algorithm can optimize any objective that is computable in polynomial time from the final value matrix $(v_i(A_j))_{i,j\in N}$, whose $(i,j)$-entry is agent $i$'s value for agent $j$'s final bundle. Examples include $\sum_i v_i(A_i)$ and $\min_i v_i(A_i)$.
\end{theorem}

Let $B:=\{b_t:t\in[T]\}$ be the set of round-end indices. A state is $s=((\Delta_{ij})_{i\ne j},(p_{ij})_{i\ne j},(q_i)_{i\in N})$.
The initial state has every entry equal to zero. Let $D_\tau$ be the set of states reachable after the first $\tau$ assignments that satisfy \eqref{eq:integer-ef1-test} at every completed round end. For each stored state, the algorithm saves a predecessor state and the recipient of the last item.

Suppose item $o$ is assigned to agent $h$.
For every $i\neq h$, update $\Delta_{ih}\gets \Delta_{ih}+v_i(o)$ and $\Delta_{hi}\gets \Delta_{hi}-v_h(o)$.
For each $i\neq h$ with $v_i(o)>0$, also set $p_{ih}\gets\max\{p_{ih},v_i(o)\}$.
If $v_h(o)<0$, set $q_h\gets\max\{q_h,-v_h(o)\}$.
All other entries remain unchanged. The dynamic program
tries every possible recipient, keeps one copy of each
resulting state, and applies~\eqref{eq:integer-ef1-test} at round ends. Full
pseudocode appears in Appendix~\ref{app:bounded-dp}.

For every $i\ne j$, $-mU\le\Delta_{ij}\le mU$ and $p_{ij},q_i\in\{0,1,\dots,U\}$.
There are $n(n-1)$ entries $\Delta_{ij}$, $n(n-1)$ entries $p_{ij}$,
and $n$ entries $q_i$. Hence the number of states is at most $S=(2mU+1)^{n(n-1)}(U+1)^{n^2}$.
For fixed $n$, the running time is
$O((mn^2+Tn^2)S)$.

The final values $v_i(O)$ and the final values of $\Delta_{ij}$ determine
the complete final value matrix. For every agent $i$, $v_i(A_i)=
\frac{v_i(O)-\sum_{j\ne i}\Delta_{ij}}{n}$ and $v_i(A_j)=v_i(A_i)+\Delta_{ij}$.
Thus the stated objective can be evaluated for every final state.

\section{Limits of Temporal Maximin Share Fairness}
\label{sec:tmms}

We now ask what maximin share guarantee follows from TEF1 and how hard it
is to decide whether an exact temporal maximin share allocation exists.
For $S\subseteq O$, define agent $i$'s maximin share by $\operatorname{MMS}_i(S):=
\max_{(P_1,\dots,P_n)\in\Pi_n(S)}
\min_{h\in N}v_i(P_h)$.
For $\alpha\in[0,1]$, an allocation $A$ is
\emph{$\alpha$-temporal maximin share fair} ($\alpha$-TMMS) for goods if $v_i(A_i^t)\ge
\alpha\operatorname{MMS}_i(O_{\le t})$ for every $i \in N$ and $t \in [T]$.
We write TMMS for $1$-TMMS.

\begin{theorem}
\label{thm:tef1-tmms}
    For additive goods, every TEF1 allocation is $1/n$-TMMS. Moreover, for
    every $n\ge2$ and every $\alpha\in(1/n,1]$, there is a two-round goods
    instance with identical valuations that has a TEF1 allocation but has no
    $\alpha$-TMMS allocation.
\end{theorem}

For the second statement, let round 1 contain $n$ goods of value $1$, and
let round 2 contain $n-1$ goods of value $n$. The round-1 MMS is $1$.
Therefore any $\alpha$-TMMS allocation with $\alpha>0$ must give every
agent positive value in round 1; because there are exactly $n$ unit goods,
every agent must receive one. At the end of round 2, the MMS is $n$, but
at least one agent receives no value-$n$ good and has final value $1$.
Hence no factor greater than $1/n$ is possible. On the other hand, giving
one unit good to every agent and the value-$n$ goods to distinct agents
produces a TEF1 allocation.

This instance also satisfies Definition~\ref{def:scaled-agreement} with
$c_i=1$ for every agent. Thus the $1/n$ bound remains best possible even
in the valuation class of Theorem~\ref{thm:scaled-allocation}.

This implication is specific to EF1 and does not extend to $\mathrm{EF}\ell$ for $\ell\geq 2$. Fix $n\ge2$ and $\ell\ge2$, and
consider a one-round instance with $n$ identical unit-valued goods. Give two
goods to one agent, one good to each of $n-2$ agents, and no good to the
remaining agent. The allocation is TEF$\ell$, but the last agent receives
value $0$ while her MMS is $1$. Thus TEF$\ell$ with $\ell\ge2$ gives no
positive TMMS factor in general.

For an instance containing only chores, define the nonnegative cost
function $d_i(o):=-v_i(o)$, $d_i(S):=\sum_{o\in S}d_i(o)$, 
and define $\mu_i(S):=
\min_{(P_1,\dots,P_n)\in\Pi_n(S)}
\max_{h\in N}d_i(P_h)$.
Exact TMMS for chores requires  $d_i(A_i^t)\le\mu_i(O_{\le t})$ for every $i \in N$ and $t \in [T]$.

\begin{theorem}
\label{thm:tmms-hardness}
Deciding whether an exact TMMS allocation exists is NP-hard, even for two
agents, two rounds, and identical valuations. The result holds for both
goods and chores.
\end{theorem}

\section{Conclusion}
We identified four cases in which temporal fair division of indivisible
mixed manna can be handled efficiently. With at most $k$ item types, an
online cyclic rule is EF$\lceil k/2\rceil$ after every item
arrival, and TEF1 existence is polynomial-time decidable for fixed $n$
and $k$. Under the agreement-after-scaling condition in
Definition~\ref{def:scaled-agreement}, given a valid scale vector before the
arrivals, an online rule is EF1 and Pareto optimal after every item arrival. A different EF1 rule applies to a
two-part common-ranking sequence. For fixed $n$ and integer values in
$[-U,U]$, an exact pseudo-polynomial dynamic program decides TEF1
existence.

For additive goods, TEF1 gives every agent at least $1/n$ of her MMS at
every round, and no larger factor follows from TEF1 in general, even with
identical valuations and two rounds. Deciding exact TMMS existence is
NP-hard for two agents and two rounds for both goods and chores.

A natural next question is to identify broader mixed manna valuation
classes that admit an online TEF1 allocation, especially together with
Pareto optimality after every item arrival.
\bibliographystyle{plainnat}
\bibliography{bib}

\newpage

\appendix

\section*{APPENDIX}

\section{Further Related Work}

\paragraph{Online and repeated allocation.}
Standard online fair division does not reveal later
items~\citep{aleksandrov2015,aleksandrov2019}. Fairness and
efficiency under worst-case or random arrivals are studied by
\citet{zeng2020fairness_efficiency_dynamicFD} and \citet{benade2024}. Other work gives online algorithms limited information about later items, such as totals, a few future values, or predictions~\citep{benade2025partial,neoh2026,
amanatidis2025personalized2value,melissourgos2026predictions}. Recent papers also study EF1, envy-freeness up to any item (EFX), MMS,
proportionality, and other approximate fairness guarantees during or at
the end of an online sequence, under restricted values or an unknown number of arrivals~\citep{zhou2023icml_mms_chores,song2025onlinechores,wang2026onlinefairBinary,choo2025approxproponline,kahana2026perpetual,chen2026competitive}. \citet{amanatidis2026buffers} are especially close to our online results for mixed manna: they obtain EF1 after every step by allowing
a bounded number of items to be kept temporarily and assigned later.
Our online rules keep no item for later and assign the current item
immediately.  

In repeated allocation models, the same set of items or positions is
assigned again in later rounds. In our model, each arriving item is
assigned once and never returns.
\citet{gollapudi2020repeated} study cumulative
EF1-style fairness in repeated two-sided matching. \citet{igarashi2023repeatedfairallication} study recurring goods and chores and
compare fairness within one round with fairness over all rounds.
\citet{caragiannis2023repeatedmatching} and \citet{micheel2024house} study repeated one-to-one or house
assignment. \citet{Lim2026repeated} maximize the least
cumulative utility, either at the end or after every round. \citet{adams2026perpetually} study ordinal proportionality (i.e., proportionality defined only from agents' rankings)
up to one or two items after every day. These models use recurring
items or one-to-one assignments, whereas a round in our model may
contain any number of new items.

\paragraph{Other temporal collective decision-making models.}
A parallel line of work studies sequences of public decisions
rather than allocations of private items. \citet{ElkindObraztsovaTeh2024TemporalFairness} give a unified
framework for temporal fairness in multiwinner voting. In a
model that selects one alternative per round, \citet{elkind2024temporal} study welfare, strategyproofness, and proportionality; subsequent work studies the verification of temporal proportionality~\citep{elkind2025verifying}, stronger
proportionality axioms~\citep{phillips2026strengthening}, and
their welfare cost~\citep{teh2026price}. Related temporal models
include assigning projects to time slots under welfare and
fairness objectives~\citep{elkind2022temporalslot}, choosing
committees while limiting changes between
stages~\citep{zech2024multiwinner}, and voting over public chores
with dynamic preferences~\citep{elkind2025nimby}. These settings
choose public outcomes or assign positions, rather than
partitioning private items among agents. Their representation
and welfare criteria are therefore not directly comparable with
TEF1, but they share the broader goal of controlling fairness
and efficiency across a sequence of decisions.

\section{Omitted Proofs from Section~\ref{sec:limited-item-diversity}}

\subsection{Proof of Theorem~\ref{thm:few-types}}
Fix the item order used by Algorithm~\ref{alg:few-types}, and use the
notation $F,R,P_r,Z_r,C_r,\sigma_r$ from Section~\ref{sec:limited-item-diversity}. After $q'$ types have
appeared, the balancing step gives
\[
\max\{|F|,|R|\}\le\left\lceil\frac{q'}{2}\right\rceil
\le\left\lceil\frac{q}{2}\right\rceil.
\]
Fix an arbitrary item prefix and let $A_i$ be agent $i$'s bundle at this
prefix. For every type $r$, let $x_{i,r}:=|A_i\cap S_r|$.

Since items of each type are allocated cyclically according to a fixed order,
for every type $r$ and every pair of agents $i,j$ we have $|x_{i,r}-x_{j,r}|\leq 1$. Moreover, if agent $i$ appears before agent $j$ in the cyclic order $\sigma_r$,
then $x_{i,r}\geq x_{j,r}$.
Indeed, within any prefix of a cyclic sequence, an earlier position in the cycle receives either the same number of items as a later position or one more.

Now fix an ordered pair of distinct agents $i,j$. We study envy from $i$ toward
$j$. A type $r$ can contribute negatively to the comparison between $i$'s own
bundle and $j$'s bundle only in one of the following two cases:
\[
w_{i,r}>0
\quad\text{and}\quad
x_{j,r}>x_{i,r},
\]
or
\[
w_{i,r}<0
\quad\text{and}\quad
x_{i,r}>x_{j,r}.
\]
In the first case, agent $j$ has one more item of a type that agent $i$ views
as a good. In the second case, agent $i$ has one more item of a type that
agent $i$ views as a chore. We call such a type \emph{unfavorable for the ordered pair}
$(i,j)$.

We first consider the case $i<j$. We claim that no type $r\in F$ is unfavorable for
$(i,j)$. Let $r\in F$.

If $w_{i,r}>0$, then $i\in P_r$. In the order $\sigma_r$ for types in $F$, the
agents in $P_r$ are placed first and are ordered increasingly. Therefore,
because $i<j$, agent $i$ appears before agent $j$ in $\sigma_r$, regardless of
whether $j$ belongs to $P_r$, $Z_r$, or $C_r$. Hence $x_{i,r}\geq x_{j,r}$,
so it is not the case that $x_{j,r}>x_{i,r}$.

If $w_{i,r}<0$, then $i\in C_r$. In the order $\sigma_r$ for types in $F$, the
agents in $C_r$ are placed last and are ordered decreasingly. Therefore,
because $i<j$, agent $j$ appears before agent $i$ in $\sigma_r$, regardless of
whether $j$ belongs to $P_r$, $Z_r$, or $C_r$. Hence $x_{j,r}\geq x_{i,r}$,
so it is not the case that $x_{i,r}>x_{j,r}$.

Thus, when $i<j$, no type in $F$ is unfavorable for $(i,j)$. Therefore all unfavorable types for $(i,j)$ lie in $R$, and so there are at most $|R|\le \left\lceil\frac{q}{2}\right\rceil$ such types.

The case $i>j$ is symmetric. We claim that no type $r\in R$ is unfavorable for
$(i,j)$. Let $r\in R$.

If $w_{i,r}>0$, then $i\in P_r$. In the order $\sigma_r$ for types in $R$, the
agents in $P_r$ are placed first and are ordered decreasingly. Therefore,
because $i>j$, agent $i$ appears before agent $j$ in $\sigma_r$, regardless of
whether $j$ belongs to $P_r$, $Z_r$, or $C_r$. Hence $x_{i,r}\geq x_{j,r}$,
so it is not the case that $x_{j,r}>x_{i,r}$.

If $w_{i,r}<0$, then $i\in C_r$. In the order $\sigma_r$ for types in $R$, the
agents in $C_r$ are placed last and are ordered increasingly. Therefore,
because $i>j$, agent $j$ appears before agent $i$ in $\sigma_r$, regardless of
whether $j$ belongs to $P_r$, $Z_r$, or $C_r$. Hence $x_{j,r}\geq x_{i,r}$,
so it is not the case that $x_{i,r}>x_{j,r}$.

Thus, when $i>j$, no type in $R$ is unfavorable for $(i,j)$. Therefore all unfavorable types for $(i,j)$ lie in $F$, and so there are at most $|F|\le \left\lceil\frac{q}{2}\right\rceil$ such types.

We now show how to eliminate envy from $i$ toward $j$ by removing at most
$\lceil q/2\rceil$ items in total. Since $q\le k$, this also gives the
claimed $\lceil k/2\rceil$ bound. Let $D_{ij}$ be the set of unfavorable types for the ordered pair $(i,j)$.
We have shown that $|D_{ij}|\le \left\lceil\frac{q}{2}\right\rceil$.
Initialize $X_i=X_j=\varnothing$.
For every unfavorable type $r\in D_{ij}$, do the following.

If $w_{i,r}>0$ and $x_{j,r}>x_{i,r}$, then, since $|x_{i,r}-x_{j,r}|\leq 1$, we have $x_{j,r}=x_{i,r}+1$. Choose one
item of type $r$ from $A_j$ and add it to $X_j$.

If $w_{i,r}<0$ and $x_{i,r}>x_{j,r}$, then, since $|x_{i,r}-x_{j,r}|\leq 1$, we have $x_{i,r}=x_{j,r}+1$. Choose one
item of type $r$ from $A_i$ and add it to $X_i$.

By construction,
\[
|X_i|+|X_j|=|D_{ij}|
\leq
\left\lceil \frac{q}{2}\right\rceil .
\]
Moreover, every item in $X_i$ has negative value for agent $i$, and every item in $X_j$ has positive value for agent $i$.

Let $x'_{i,r}$ and $x'_{j,r}$ denote the numbers of type-$r$ items held by
agents $i$ and $j$, respectively, after these removals. We claim that for
every type $r$, $w_{i,r}(x'_{i,r}-x'_{j,r})\geq 0$.
Indeed, if $w_{i,r}>0$, then either $r$ was not unfavorable, in which case
$x_{i,r}\geq x_{j,r}$ already, or $r$ was unfavorable and we removed one type-$r$ item
from $A_j$, making $x'_{i,r}\geq x'_{j,r}$. Hence $w_{i,r}(x'_{i,r}-x'_{j,r})\geq 0$.
If $w_{i,r}<0$, then either $r$ was not unfavorable, in which case
$x_{i,r}\leq x_{j,r}$ already, or $r$ was unfavorable and we removed one type-$r$ item
from $A_i$, making $x'_{i,r}\leq x'_{j,r}$. Hence again $w_{i,r}(x'_{i,r}-x'_{j,r})\geq 0$.
Finally, if $w_{i,r}=0$, then the contribution of type $r$ is zero.

Using additivity and the fact that all items of a given type have the same
value for agent $i$, we obtain
\[
\begin{aligned}
v_i(A_i\setminus X_i)-v_i(A_j\setminus X_j)
&=
\sum_{r=1}^q w_{i,r}(x'_{i,r}-x'_{j,r}) \\
&\geq 0.
\end{aligned}
\]
Therefore
\[
v_i(A_i\setminus X_i)\geq v_i(A_j\setminus X_j).
\]
Thus envy from $i$ toward $j$ can be eliminated by removing at most
$\lceil k/2\rceil$ items, where the removed items are only chores from $i$'s
own bundle and goods from $j$'s bundle, from $i$'s perspective.

Since the item prefix and the ordered pair $(i,j)$ were arbitrary,
the allocation is EF$\lceil q/2\rceil$ after every item
prefix and therefore TEF$\lceil q/2\rceil$. Since
$q\le k$, it is also EF$\lceil k/2\rceil$ after every
item prefix.

Each assignment uses only the current item's valuation vector, the type
orders already fixed, and the current index $z_r$ for that type. It does
not use later items, so the rule is online.

\subsection{Proof of Theorem~\ref{thm:few-types-dp}} \label{app:few-types-dp}

Algorithm~\ref{alg:few-types-dp} gives the full dynamic program.

\begin{algorithm}[t]
\caption{TEF1 decision with few item types}
\label{alg:few-types-dp}
\begin{algorithmic}[1]
\STATE $D_0\gets\{\mathbf{0}\}$ and initialize an empty predecessor table
\FOR{$\tau=0,\dots,m-1$}
    \STATE $D_{\tau+1}\gets\varnothing$
    \STATE let $r$ be the type of $o_{\tau+1}$
    \FOR{each $x\in D_\tau$}
        \FOR{each recipient $h\in N$}
            \STATE copy $x$ to $y$ and set $y_{h,r}\gets y_{h,r}+1$
            \IF{$y\notin D_{\tau+1}$}
                \STATE add $y$ to $D_{\tau+1}$ and save predecessor
                $(x,h)$
            \ENDIF
        \ENDFOR
    \ENDFOR
    \IF{$\tau+1\in B$}
        \FOR{each $y\in D_{\tau+1}$}
            \IF{$\Delta_{ij}(y)>\eta_{ij}(y)$ for some $i\ne j$}
                \STATE remove $y$ from $D_{\tau+1}$
            \ENDIF
        \ENDFOR
    \ENDIF
\ENDFOR
\IF{$D_m=\varnothing$}
    \RETURN No
\ENDIF
\STATE choose $x\in D_m$, follow the saved predecessors, and return the
recovered allocation
\end{algorithmic}
\end{algorithm}

Refine the arrival process into an arbitrary item-by-item sequence
$o_1,\dots,o_m$ that respects the order of the rounds, where $m=|O|$.
That is, all items of round $t$ appear before all items of round $t+1$.
Let $b_t := \sum_{s=1}^t |O_s|$ denote the index of the last item that has arrived by the end of round $t$.

For $\tau\in\{0,\dots,m\}$ and each type $r\in[q]$, let
\[
p_r(\tau):=|\{a\in[\tau]:o_a\in S_r\}|
\]
be the number of processed items of type $r$. A state after
$\tau$ processed items is a type-count matrix $x=(x_{i,r})_{i\in N,\ r\in[q]}$, where $x_{i,r}$ is the number of type-$r$ items assigned to
agent $i$, and $\sum_{i\in N}x_{i,r}=p_r(\tau)$ for every $r\in[q]$.
Since the count of the last agent is determined by the counts of the first $n-1$
agents, the number of possible states at any time $\tau$ is at most
\[
    \prod_{r=1}^q (p_r(\tau)+1)^{n-1}
\le
(m+1)^{q(n-1)}
\le
(m+1)^{k(n-1)}.
\]

For a state $x$, define the value that agent $i$ assigns to agent $h$'s bundle by $V_i^x(h):=\sum_{r=1}^{q}x_{h,r}w_{i,r}$.
We now describe how to check whether the allocation represented by $x$ is EF1.
For each ordered pair of agents $i,j$, define the envy gap $\Delta_{ij}(x):=V_i^x(j)-V_i^x(i)$.
If $\Delta_{ij}(x)\le 0$, then agent $i$ does not envy agent $j$. Otherwise, EF1 for this pair holds if and only if at least one of the following conditions
holds:
\[
    \exists r\in[q] \text{ such that }
x_{j,r}>0,\ w_{i,r}>0,\ \Delta_{ij}(x)\le w_{i,r},
\]
or
\[
    \exists r\in[q] \text{ such that }
x_{i,r}>0,\ w_{i,r}<0,\ \Delta_{ij}(x)\le -w_{i,r}.
\]
Indeed, the first condition says that removing one good, from agent $i$'s perspective,
from agent $j$'s bundle eliminates $i$'s envy. The second condition says that removing one weak chore from
agent $i$'s own bundle eliminates $i$'s envy. Since all items of the same type have the same value for every agent, it is enough to check the $q$ types. Thus, EF1 of a state can be
verified in $O(n^2q)$ time.

Algorithm~\ref{alg:few-types-dp} applies these transitions to every current
state and filters the resulting states exactly at the round-end indices
$b_t$. We now prove correctness and bound its running time.

The correctness follows by induction on $\tau$. The transition enumerates every
possible irrevocable assignment of the next item. Since we only filter at indices
$b_t$, the dynamic program enforces EF1 exactly after every round. Therefore $D_m$ is nonempty if and only if a TEF1 allocation exists.
Following the saved predecessors returns the corresponding sequence of
recipients and hence the allocation.

At most $(m+1)^{q(n-1)}$ states are stored after any item. Each state has
$n$ possible next recipients. This gives $O(mn(m+1)^{q(n-1)})$ transitions. At each of the $T$ round ends, the EF1 test takes
$O(n^2q)$ time per state. The total running time is therefore
\[
O((mn+Tn^2q)(m+1)^{q(n-1)}),
\]
which is polynomial for fixed $n$ and $k$, because $q\le k$.

\section{Omitted Proofs from Section~\ref{sec:scaled-agreement}} 
\subsection{Finding a Valid Scale Vector}\label{app:find-scales}

For rational valuations,
Algorithm~\ref{alg:find-scales} returns a valid scale vector
when one exists and returns \textsc{No} otherwise.

\begin{algorithm}[t]
\caption{Finding a valid scale vector for rational valuations}
\label{alg:find-scales}
\begin{algorithmic}[1]
\STATE $E\gets\varnothing$
\FOR{each item $o$}
    \STATE $P_o^+\gets\{i\in N:v_i(o)>0\}$
    \IF{$P_o^+\ne\varnothing$}
        \FOR{each pair $i<j$ in $P_o^+$}
            \STATE add the equation
            $c_i/c_j=v_i(o)/v_j(o)$ to $E$
        \ENDFOR
    \ELSIF{$v_i(o)<0$ for every $i\in N$}
        \FOR{each pair $i<j$ in $N$}
            \STATE add the equation
            $c_i/c_j=v_i(o)/v_j(o)$ to $E$
        \ENDFOR
    \ENDIF
\ENDFOR
\STATE form a graph $H$ with one vertex for each agent and one edge for
      each equation in $E$
\STATE leave every $c_i$ undefined and mark every equation unchecked
\FOR{each connected component $K$ of $H$}
    \STATE choose the smallest-label agent $r\in K$, set $c_r\gets1$,
          and place $r$ in a queue
    \WHILE{the queue is nonempty}
        \STATE remove an agent $u$ from the queue
        \FOR{each unchecked equation $c_i/c_j=\rho$ incident to $u$}
            \STATE mark the equation checked
            \IF{$c_i$ is undefined}
                \STATE $c_i\gets\rho c_j$ and place $i$ in the queue
            \ELSIF{$c_j$ is undefined}
                \STATE $c_j\gets c_i/\rho$ and place $j$ in the queue
            \ELSIF{$c_i/c_j\ne\rho$}
                \RETURN No
            \ENDIF
        \ENDFOR
    \ENDWHILE
\ENDFOR
\RETURN $(c_1,\dots,c_n)$
\end{algorithmic}
\end{algorithm}

All ratios used by the algorithm are positive.
The equations are exactly those required by
Definition~\ref{def:scaled-agreement}. Hence a valid scale vector exists exactly
when these equations are consistent. The algorithm uses
$O(mn^2)$ arithmetic operations and comparisons.

\subsection{Proof of Theorem~\ref{thm:scaled-allocation}}
Work with the normalized additive valuations $a_1,\ldots,a_n$ from Definition~4.1. Because $c_i>0$, for all sets $X,Y\subseteq O$,
\[
v_i(X)\geq v_i(Y) \quad\Longleftrightarrow\quad a_i(X)\geq a_i(Y).
\]
Moreover, $v_i(o)$ and $a_i(o)$ have the same sign for every item $o$. Therefore, an allocation is EF1 under the original valuations if and only if it is EF1 under the normalized valuations.

Algorithm~\ref{alg:scaled-allocation} assigns each item to an agent whose
normalized value for that item is largest. Indeed, in case (i) of
Definition~\ref{def:scaled-agreement}, the recipient has value
$p_o=\max_i a_i(o)$; in case (ii), the recipient has value
$0=\max_i a_i(o)$; and in case (iii), every agent has value $q_o$. Hence, at
any time, if $x\in A_j$, then $a_i(x)\le a_j(x)$ for every $i \in N$.
Summing over $x\in A_j$ gives
\begin{equation}
\label{eq:assigned-item-max}
a_i(A_j)\le a_j(A_j)=L_j
\quad\text{for all }i,j\in N.
\end{equation}

We prove by induction on $s\in\{0,\dots,m\}$ that the allocation of
$Q_s$ is EF1. The claim is immediate for $s=0$. Fix $s\ge1$, let
$o:=o_s$, and suppose the allocation $A$ of $Q_{s-1}$ is EF1. Let $h$ be
the recipient chosen for $o$, and let $A'$ be the new allocation.

First suppose $P_o\ne\varnothing$. Let $i\in P_o\setminus\{h\}$. By~\eqref{eq:assigned-item-max} and the choice of $h$,
\[
a_i(A_h)\le L_h\le L_i=a_i(A_i).
\]
Since $a_i(o)=p_o>0$, $a_i(A'_i)\ge a_i(A'_h\setminus\{o\})$.
Thus deleting $o$ from $h$'s bundle removes any envy of $i$ toward $h$.
If $i\notin P_o$, then $a_i(o)\le0$, so $a_i(A'_h)\le a_i(A_h)$.
Any deletion that established EF1 for the comparison of $i$ with $h$ before the assignment remains valid afterward. Comparisons from $h$ toward other
agents also remain valid because $h$'s own normalized bundle value has
increased by $p_o>0$. All other ordered pairs are unchanged. Hence $A'$
is EF1 in this case.

Next suppose $P_o=\varnothing$ and $\max_i a_i(o)=0$. The chosen agent
$h$ has $a_h(o)=0$, while $a_i(o)\le0$ for every agent $i$. Therefore
$h$'s own bundle value is unchanged, and every other agent weakly
decreases her value for $h$'s bundle. All EF1 comparisons that held under
$A$ continue to hold under $A'$.

Finally, suppose $a_i(o)=q_o<0$ for every agent $i$. Let $j\ne h$. By
Equation~\eqref{eq:assigned-item-max} and the choice of $h$ as an agent
with maximum current normalized bundle value,
\[
a_h(A_j)\le L_j\le L_h=a_h(A_h).
\]
Since $A'_h\setminus\{o\}=A_h$, deleting $o$ from $h$'s own bundle gives
\[
a_h(A'_h\setminus\{o\})\ge a_h(A'_j).
\]
Thus the EF1 comparison from $h$ toward every other agent holds. For any
$i\ne h$, the value $a_i(A'_h)$ is smaller than $a_i(A_h)$ because
$a_i(o)=q_o<0$. Hence every comparison from another agent toward $h$
becomes weakly easier, and all other ordered pairs are unchanged. Thus
$A'$ is EF1.

The induction proves EF1 after every item arrival. Since each $c_i$ is positive, normalization does not change comparisons or whether a value is positive, zero, or negative. Therefore the same conclusion holds under the original valuations.

We next prove the weighted-sum claim. Fix $s$ and let $A$ be the
allocation of $Q_s$ returned by Algorithm~\ref{alg:scaled-allocation}.
Every item is assigned to an agent with largest normalized value, so
\[
\sum_{i\in N}a_i(A_i)
=
\sum_{o\in Q_s}\max_{k\in N}a_k(o).
\]
For any other allocation $B\in\Pi_n(Q_s)$,
\[
\sum_{i\in N}a_i(B_i)
=
\sum_{i\in N}\sum_{o\in B_i}a_i(o)
\le
\sum_{o\in Q_s}\max_{k\in N}a_k(o).
\]
Therefore
\[
\sum_{i\in N}\frac{v_i(A_i)}{c_i}
=
\max_{B\in\Pi_n(Q_s)}
\sum_{i\in N}\frac{v_i(B_i)}{c_i}.
\]
If another allocation Pareto dominated $A$, every term in this weighted
sum would weakly increase and at least one would strictly increase,
because every $c_i$ is positive. This would contradict the displayed
maximum. Hence the allocation is Pareto optimal.

At the arrival of item $o_s$, the rule uses only the normalized values of
$o_s$, the current values $L_1,\dots,L_n$, the supplied scales, and the
fixed smallest-label tie rule. It does not use later items, so it is
online. One scan of the $n$ agents identifies the applicable case and the
recipient. The total running time is $O(mn)$.

\section{Omitted Proofs from Section~\ref{sec:common-ranking}}

\subsection{Finding a Valid Split and Set $P_G$}
\label{app:find-split}

Algorithm~\ref{alg:find-common-ranking} tries every split index and
returns a valid choice when one exists.

\begin{algorithm}[t]
\caption{Finding a valid split and set $P_G$}
\label{alg:find-common-ranking}
\begin{algorithmic}[1]
\FOR{$b=0,\dots,m$}
    \STATE $C_b\gets\{o_1,\dots,o_b\}$ and
    $G_b\gets\{o_{b+1},\dots,o_m\}$
    \IF{$C_b$ does not satisfy condition 1 of
    Definition~\ref{def:common-ranking}}
        \STATE continue
    \ENDIF
    \IF{$G_b=\varnothing$}
        \RETURN $(C_b,G_b,\varnothing)$
    \ENDIF
    \STATE $P^+\gets\{i\in N:v_i(o)>0\text{ for some }o\in G_b\}$
    \IF{$P^+\ne\varnothing$}
        \IF{$G_b$ satisfies condition (ii) with $P_G=P^+$}
            \RETURN $(C_b,G_b,P^+)$
        \ENDIF
    \ELSIF{some $i\in N$ has $v_i(o)=0$ for every $o\in G_b$}
        \RETURN $(C_b,G_b,\{i\})$
    \ENDIF
\ENDFOR
\RETURN No
\end{algorithmic}
\end{algorithm}

To see that the search is complete, first observe that every agent who assigns positive value to some item of $G_b$ must belong to every valid $P_G$. Hence, if $P^+\ne\varnothing$ and a valid set exists, then $P^+$
itself satisfies condition~(ii): its agents satisfy the nonnegativity and
ranking requirements, while every agent outside $P^+$ values every item of
$G_b$ nonpositively. If $P^+=\varnothing$, a nonempty valid set can contain
only agents who value every item of $G_b$ at zero, and any such singleton is
valid. For each split, the conditions can be checked in $O(mn)$ time, so the
total running time is $O(m^2n)$.

\subsection{Proof of Lemma~\ref{lem:chore-cycle}}

If $C=\varnothing$, the claim is immediate. Otherwise, let
$M:=d+|C|$. For the proof only, place $d$ additional zero-valued items
before the items of $C$, and call the resulting ordered list
$x_1,\dots,x_M$. These additional items do not belong to the instance.
For every agent $i$,
\[
v_i(x_1)\ge v_i(x_2)\ge\cdots\ge v_i(x_M),
\]
and every value in the list is nonpositive. Allocate this list
cyclically in the order $1,2,\dots,n$.

First consider an arbitrary prefix of the padded sequence. Fix an
ordered pair of agents $i,j$, and evaluate all items using
$v_i$. Let $p_1<\cdots<p_r$ be the positions in this prefix assigned to agent $i$, and let $q_1<\cdots<q_s$ be the positions assigned to agent $j$.

Suppose first that $i<j$. Then $r=s$ or $r=s+1$, and for
every $a\leq s$ we have $p_a<q_a$. Hence
\[
v_i(x_{p_a})\ge v_i(x_{q_a}) \quad\text{for every } a\leq s.
\]
If $r=s$, agent $i$ does not envy agent $j$. If $r=s+1$,
then deleting the last item $x_{p_r}$ from agent $i$'s bundle
leaves a bundle whose value is at least the value of agent $j$'s
bundle. The deleted item is a weak chore for agent $i$.

Now suppose that $i>j$. Then $s=r$ or $s=r+1$. If $s=r+1$,
then for every $a\leq r$ we have $p_a<q_{a+1}$, and therefore
\[
\sum_{a=1}^r v_i(x_{p_a})
   \geq
\sum_{a=2}^{s} v_i(x_{q_a})
   \geq
\sum_{a=1}^{s} v_i(x_{q_a}),
\]
where the last inequality uses $v_i(x_{q_1})\leq 0$. Thus agent
$i$ does not envy agent $j$. If $s=r=0$, neither $i$ nor $j$ has received an item, so
there is no envy. Suppose now that $s=r\ge 1$. After deleting
the last item $x_{p_r}$ from agent $i$'s bundle, the same
pairing gives
\[
\sum_{a=1}^{r-1}v_i(x_{p_a})
\ge
\sum_{a=2}^{s}v_i(x_{q_a})
\ge
\sum_{a=1}^{s}v_i(x_{q_a}).
\]
Again, the deleted item is a weak chore for agent $i$.

Thus every prefix of the padded allocation is EF1. Deleting
dummy zero-valued items does not change any bundle value. If the argument above deletes a dummy item, then the same inequality
already holds without deleting a real item. Hence every
real prefix of $C$ is EF1.

It remains to prove the directional statement after the full part.
Since $M$ is divisible by $n$, every agent receives the same
number of items in the padded allocation. If $i<j$, then the
positions assigned to $i$ are earlier cycle-by-cycle than the
positions assigned to $j$. Since the sequence is ranked for agent
$i$, agent $i$ does not envy agent $j$. Conversely, for envy
of $j$ toward $i$, we are in the case $j>i$ with equal bundle
sizes. The argument above shows that deleting the last padded item
from $j$'s own bundle removes the envy. If that item is a dummy
zero-valued item, no deletion is needed in the real allocation; otherwise
it is a weak chore in $j$'s real bundle. This proves the claim.

\subsection{Proof of Theorem~\ref{thm:common-ranking}}

If $C\ne\varnothing$, allocate it according to
Lemma~\ref{lem:chore-cycle}. If $G\ne\varnothing$, denote $P_G=\{p_1<\cdots<p_g\}$
and allocate the items of $G$ cyclically in the order
\[
p_g,p_{g-1},\dots,p_1.
\]
Agents outside $P_G$ receive no item from $G$.

Every prefix that ends within $C$ is EF1 by
Lemma~\ref{lem:chore-cycle}. If $G=\varnothing$, this completes
the proof. We therefore consider a prefix containing all of $C$
and some prefix of $G$.

Fix two agents $i<j$. On $C$, agent $i$ does not envy agent
$j$, either by Lemma~\ref{lem:chore-cycle} or trivially when
$C=\varnothing$.

We first compare their bundles from $G$ using $v_i$. If at least one of $i$ and $j$ lies outside $P_G$, the $G$-items create no envy from $i$ toward $j$. If $i\in P_G$ and $j\notin P_G$, then $i$ receives weak goods and $j$ receives no $G$-items. Otherwise, $i\notin P_G$, so $i$ assigns nonpositive value to every $G$-item that $j$ receives.

Now suppose $i,j\in P_G$. Since $i<j$, agent $j$ appears before agent
$i$ in the decreasing cyclic order on $P_G$. Let
$y_1,\dots,y_u$ be the items of the current part of $G$ assigned to
$j$, and let $z_1,\dots,z_r$ be those assigned to $i$, each listed in
arrival order. Then $u\in\{r,r+1\}$, and $z_a$ arrives before
$y_{a+1}$ whenever both items exist. If $u=0$, then $r=0$, so neither agent has received an item from $G$ and
there is nothing to prove. If $r=0$, then $u=1$, and deleting the weak good
$y_1$ leaves both agents with an empty $G$-bundle. Hence assume $r\ge1$.
Since agent $i$ weakly prefers earlier items of $G$,
\[
v_i(z_a)\ge v_i(y_{a+1})
\quad\text{for }a=1,\dots,r-1.
\]
If $u=r+1$, the same inequality also holds for $a=r$, and pairing
$z_a$ with $y_{a+1}$ for $a=1,\dots,r$ gives
\[
v_i(A_i\cap G)\ge v_i((A_j\cap G)\setminus\{y_1\}).
\]
If $u=r$, pair $z_1,\dots,z_{r-1}$ with $y_2,\dots,y_r$; the remaining
item $z_r$ has nonnegative value for $i$, so the same inequality holds.
Thus deleting the weak good $y_1$ from $j$'s bundle removes any envy of $i$
toward $j$ created by the items of $G$.

For the reverse direction, consider agent $j$'s comparison with
$i$. On $G$, agent $j$ does not envy agent $i$. If both
agents belong to $P_G$, the $a$-th item assigned to $j$
arrives no later than the $a$-th item assigned to $i$, and
agent $j$ receives at least as many items. Since agent $j$'s
values along $G$ are nonincreasing and all these items are weak
goods for $j$, $v_j(A_j\cap G)\ge v_j(A_i\cap G)$.
If at least one of the two agents is outside $P_G$, the same
inequality follows directly from the nonnegative and nonpositive value conditions.

On $C$, Lemma~\ref{lem:chore-cycle} says that any envy of $j$
toward $i$ can be removed by deleting one weak chore from $j$'s
own bundle; when $C=\varnothing$, there is no envy on this part.
Together with the no-envy comparison on $G$, the same deletion
removes envy in the full prefix.

The argument holds for every item prefix and every ordered pair of
agents. Hence the allocation is EF1 after every item prefix, and
therefore TEF1. The rule can be implemented in polynomial time.

\subsection{The Two Valuation Classes Are Incomparable}
\label{app:class-comparison}
First, consider three agents and two items:
\[
\begin{array}{c|cc}
 & g_1 & g_2\\ \hline
v_1 & 4 & 3\\
v_2 & 100 & 1\\
v_3 & -1 & -2
\end{array}
\]
The two items form the $G$ part of a two-part common-ranking sequence with
$P_G=\{1,2\}$. However, no positive scales $c_1,c_2$ satisfy
Definition~\ref{def:scaled-agreement}, because equality for both items
would require $\frac{4}{c_1}=\frac{100}{c_2}$ and $\frac{3}{c_1}=\frac{1}{c_2}$,
which cannot both hold.

Conversely, consider two agents and two items arriving in the displayed
order:
\[
\begin{array}{c|cc}
 & g_1 & g_2\\ \hline
v_1 & 1 & -1\\
v_2 & -1 & 1
\end{array}
\]
With $c_1=c_2=1$, the profile satisfies
Definition~\ref{def:scaled-agreement}. It does not satisfy
Definition~\ref{def:common-ranking}: neither item can be placed in $C$,
and no fixed set $P_G$ assigns nonnegative value to both items. Hence
neither class contains the other.
\section{Omitted Proofs from Section~\ref{sec:integer-values}}

\subsection{Proof of Theorem~\ref{thm:integer-dp}} \label{app:bounded-dp}

Algorithm~\ref{alg:integer-dp} gives the full dynamic program.

\begin{algorithm}[t]
\caption{TEF1 decision with bounded integer values}
\label{alg:integer-dp}
\begin{algorithmic}[1]
\STATE $D_0\gets\{s_0\}$, where every entry of $s_0$ is zero
\STATE initialize an empty predecessor table
\FOR{$\tau=0,\dots,m-1$}
    \STATE $D_{\tau+1}\gets\varnothing$ and $o\gets o_{\tau+1}$
    \FOR{each $s\in D_\tau$}
        \FOR{each recipient $h\in N$}
            \STATE copy $s$ to $s'$
            \FOR{each $i\in N\setminus\{h\}$}
                \STATE $\Delta'_{ih}\gets\Delta_{ih}+v_i(o)$
                \STATE $\Delta'_{hi}\gets\Delta_{hi}-v_h(o)$
                \IF{$v_i(o)>0$}
                    \STATE $p'_{ih}\gets\max\{p_{ih},v_i(o)\}$
                \ENDIF
            \ENDFOR
            \IF{$v_h(o)<0$}
                \STATE $q'_h\gets\max\{q_h,-v_h(o)\}$
            \ENDIF
            \IF{$s'\notin D_{\tau+1}$}
                \STATE add $s'$ to $D_{\tau+1}$ and save predecessor
                $(s,h)$
            \ENDIF
        \ENDFOR
    \ENDFOR
    \IF{$\tau+1\in B$}
        \FOR{each $s'\in D_{\tau+1}$}
            \IF{$\Delta'_{ij}>\max\{p'_{ij},q'_i\}$ for some $i\ne j$}
                \STATE remove $s'$ from $D_{\tau+1}$
            \ENDIF
        \ENDFOR
    \ENDIF
\ENDFOR
\IF{$D_m=\varnothing$}
    \RETURN No
\ENDIF
\STATE choose the best final state for the given objective, or any final
state if no objective is specified
\STATE follow the saved predecessors and return the recovered allocation
\end{algorithmic}
\end{algorithm}

Fix an arbitrary item-by-item order $o_1,\ldots,o_m$ that respects the round order; that is, every item in round $t$ appears before every item in round $t+1$. Let $b_t:=\sum_{s=1}^t |O_s|$ denote the index of the last item that has arrived by the end of round $t$.

For a partial allocation, use the quantities
$\Delta_{ij}$, $p_{ij}$, and $q_i$ defined in
Section~\ref{sec:integer-values}. As shown in
\eqref{eq:integer-ef1-test}, the allocation is EF1 if and only if $\Delta_{ij}\le \max\{p_{ij},q_i\}$ for every $i \neq j$.

A state is $s=
\left(
  (\Delta_{ij})_{i\ne j},
  (p_{ij})_{i\ne j},
  (q_i)_{i\in N}
\right)$.
The initial state has every entry equal to zero.

Suppose that item $o$ is assigned to agent $h$. For every ordered
pair $i\ne j$, update the envy gap by
\[
\Delta'_{ij}
=
\begin{cases}
\Delta_{ij}+v_i(o), & \text{if }j=h,\\
\Delta_{ij}-v_i(o), & \text{if }i=h,\\
\Delta_{ij},        & \text{otherwise}.
\end{cases}
\]
The two nontrivial cases cannot occur simultaneously because
$i\ne j$.

For every ordered pair $i\ne j$, update
\[
p'_{ij}
=
\begin{cases}
\max\{p_{ij},v_i(o)\},
  & \text{if }j=h\text{ and }v_i(o)>0,\\
p_{ij}, & \text{otherwise},
\end{cases}
\]
and, for every $i\in N$, update
\[
q'_i
=
\begin{cases}
\max\{q_i,-v_i(o)\},
  & \text{if }i=h\text{ and }v_i(o)<0,\\
q_i, & \text{otherwise}.
\end{cases}
\]
These formulas determine the next state using only the current state,
the next item, and its recipient.

Algorithm~\ref{alg:integer-dp} starts from the all-zero state, tries every
recipient for each item using the updates above, and filters states by
\eqref{eq:integer-ef1-test} exactly at the round-end indices. We now prove
correctness and bound its running time.

Every state retained at a round end corresponds to an allocation that is EF1 at that round and at every earlier round. Conversely, fix any TEF1 allocation and follow its recipient choices item by item. The corresponding state is generated at every step, and none of its round-end states is removed. Therefore $D_m$ is nonempty exactly when a TEF1
allocation exists. Following the saved predecessors returns one such
allocation.

It remains to bound the number of states. For every $i\ne j$, $-mU\le \Delta_{ij}\le mU$ because each item contributes at most $U$ in absolute value to the
difference $v_i(A_j)-v_i(A_i)$. Also, $p_{ij},q_i\in\{0,1,\dots,U\}$.
There are $n(n-1)$ envy-gap entries, $n(n-1)$ entries $p_{ij}$,
and $n$ entries $q_i$. Hence the number of states is at most
\[
S
=
(2mU+1)^{n(n-1)}
(U+1)^{n^2}.
\]

For a fixed recipient, one transition changes $O(n)$ entries.
Trying all $n$ recipients therefore takes $O(n^2)$ time per state
and item. Checking \eqref{eq:integer-ef1-test} takes $O(n^2)$ time per
state at each round end. The total running time is $O\left( (mn^2+Tn^2)S \right)$, which is pseudo-polynomial for fixed $n$.

Finally, the envy gaps determine the final value matrix. For every
agent $i$,
\[
v_i(O)
=
\sum_{h\in N}v_i(A_h)
=
n\,v_i(A_i)+\sum_{j\ne i}\Delta_{ij}.
\]
Therefore
\[
v_i(A_i)
=
\frac{
  v_i(O)-\sum_{j\ne i}\Delta_{ij}
}{n},
\]
and then
\[
v_i(A_j)=v_i(A_i)+\Delta_{ij}
\quad\text{for }j\ne i.
\]
Thus every objective determined by the final value matrix can be evaluated
on each reachable final state, and a best final state can be selected.

\section{Omitted Proofs from Section~\ref{sec:tmms}}

\subsection{Proof of Theorem~\ref{thm:tef1-tmms}}

Since all item values are nonnegative, if agent $i$ strictly envies
agent $j$, deleting a weak chore from $i$'s own bundle cannot reduce the
envy; EF1 therefore provides an item in $j$'s bundle whose deletion
removes it.

First fix a round $t$ and an agent $i$, and let
$x:=v_i(A_i^t)$. For every agent $j\neq i$ such that
$v_i(A_j^t)>x$, EF1 gives an item $g_j\in A_j^t$ with $v_i(A_j^t\setminus\{g_j\})\le x$.
Let $R_i$ be the set of these selected items. Then
$\lvert R_i\rvert\le n-1$. For every agent $j$ whom $i$ does not envy, we already have $v_i(A_j^t)\leq x$. Hence $v_i(O_{\le t}\setminus R_i)\le nx$.

In every partition of $O_{\le t}$ into $n$ bundles, at least one bundle is disjoint from $R_i$. Since all item values are nonnegative, that bundle has
value at most $v_i(O_{\le t}\setminus R_i)\le nx$. Therefore
\[
    \operatorname{MMS}_i(O_{\le t})\le nx,
\]
and so $v_i(A_i^t)\ge \operatorname{MMS}_i(O_{\le t})/n$. Since $i$ and
$t$ were arbitrary, every TEF1 allocation is $1/n$-TMMS.

To prove that the factor cannot be improved, fix $n\ge 2$. In round~1, let
$n$ goods of value $1$ arrive. In round~2, let $n-1$ goods of value
$n$ arrive. All agents have the same valuation. The round-1 MMS is $1$.
Thus any $\alpha$-TMMS allocation with $\alpha>0$ must give every agent one
round-1 good. At the end of round~2, the MMS is $n$: place each value-$n$
good alone and place all $n$ unit goods together. At least one agent receives
no value-$n$ good, so her final value is $1<\alpha n$ whenever
$\alpha>1/n$. Hence no such allocation is $\alpha$-TMMS.

Finally, give one unit good to every agent in round~1 and give the $n-1$
value-$n$ goods to distinct agents in round~2. The round-1 allocation is
envy-free. In the final allocation, the agent without a value-$n$ good can remove the value-$n$ good from any bundle she envies, after which both compared bundles have value $1$. Thus this allocation is TEF1.

\subsection{Proof of Theorem~\ref{thm:tmms-hardness}}

We reduce from \textsc{Partition}. Given positive integers
$a_1,\dots,a_r$, the problem asks whether there is a set
$P\subseteq[r]$ such that
\[
\sum_{j\in P}a_j
=
\frac12\sum_{j=1}^r a_j.
\]
Set $s_j:=2a_j$ for every $j \in [r]$,
and let
\[
W:=\frac12\sum_{j=1}^r s_j=\sum_{j=1}^r a_j.
\]
Then a set $P$ satisfies $\sum_{j\in P}s_j=W$ exactly when the original
integers can be divided into two equal-sum parts. We may assume $r\ge2$.

\paragraph{Goods.}
Construct a two-agent, two-round instance with identical valuations. In
round 1, let
\[
O_1=\{g_1,\dots,g_r\},
\quad
v(g_j)=s_j.
\]
In round 2, let
\[
O_2=\{h_1,h_2\},
\quad
v(h_1)=v(h_2)=W.
\]

Suppose first that there is a set $P\subseteq[r]$ with
$\sum_{j\in P}s_j=W$. In round 1, give $\{g_j:j\in P\}$ to agent 1 and
the remaining round-1 goods to agent 2. Each agent receives value $W$.
The round-1 MMS is exactly $W$: the displayed allocation shows that the
MMS is at least $W$, while the smaller bundle in any two-part partition
cannot have value greater than half of the total value $2W$. In round 2,
give one value-$W$ good to each agent. Each agent then has value $2W$.
The final MMS is exactly $2W$: the two sets $O_1$ and $O_2$ each have
value $2W$, and no two-part partition can have both bundles above half of
the total value $4W$. Hence the temporal instance has an exact TMMS
allocation.

Conversely, suppose the temporal goods instance has an exact TMMS
allocation. At the end of round 2, the total value is $4W$, and the
partition $(O_1,O_2)$ shows that the final MMS is $2W$. Therefore each
agent must have final value at least $2W$, and since the total is $4W$,
each has final value exactly $2W$.

Let $x_i:=v(A_i^1)$ be agent $i$'s round-1 value, and let
$y_i:=v(A_i\cap O_2)$ be her value from round 2. Since
$y_i\in\{0,W,2W\}$ and $x_i+y_i=2W$, $x_i\in\{0,W,2W\}$.
The round-1 MMS is positive: because $r\ge2$ and every $s_j$ is positive,
one can place one round-1 good in one bundle and all remaining round-1
goods in the other, giving both bundles positive value. Exact TMMS after
round 1 therefore requires $x_i>0$ for both agents. Since
$x_1+x_2=2W$, it follows that $x_1=x_2=W$. Either agent's round-1 bundle
therefore gives a solution to \textsc{Partition}.

\paragraph{Chores.}
Use the same numbers as costs. In round 1, let
\[
O_1=\{c_1,\dots,c_r\},
\quad
d(c_j)=s_j,
\]
and in round 2 let
\[
O_2=\{e_1,e_2\},
\quad
d(e_1)=d(e_2)=W.
\]

If there is a set $P$ with $\sum_{j\in P}s_j=W$, divide the round-1
chores into two bundles of cost $W$ and give one round-2 chore to each
agent. The round-1 chore MMS is $W$: the displayed division gives a
maximum bundle cost of $W$, while every two-part partition has a bundle
of cost at least half of the total cost $2W$. The final chore MMS is
$2W$ by the same argument, using the partition $(O_1,O_2)$. Thus this
allocation is exact TMMS.

Conversely, suppose an exact TMMS allocation exists. The final total cost
is $4W$, and the final chore MMS is $2W$, so both agents must have final
cost exactly $2W$. Let $x_i:=d(A_i^1)$ and
$y_i:=d(A_i\cap O_2)$. Again,
\[
y_i\in\{0,W,2W\},
\quad
x_i+y_i=2W,
\quad
x_i\in\{0,W,2W\}.
\]
The round-1 chore MMS is strictly less than $2W$. Indeed, place one
round-1 chore in one bundle and all remaining round-1 chores in the other.
Since $r\ge2$ and every cost is positive, both bundle costs are
strictly less than the total cost $2W$. Exact TMMS after round 1 therefore
requires $x_i<2W$ for both agents. Since $x_1+x_2=2W$, it follows that
$x_1=x_2=W$. Either agent's round-1 bundle gives a solution to
\textsc{Partition}. The reduction takes polynomial time. Since it is from
\textsc{Partition}, it establishes NP-hardness but not strong NP-hardness.

\end{document}